# Lightweight Deep Autoencoder for ECG Denoising with Morphology Preservation and Near Real-Time Hardware Deployment

[1]Mahdi Pirayesh Shirazi Nejad, [2]David Hicks, [2]Matt Valentine, [1]Ki H. Chon

[1]University of Connecticut, Department of BME, Storrs, CT
[2]Defibtech, Guilford, CT

## Abstract

Electrocardiogram (ECG) signals are often degraded by various noise sources such as baseline wander, motion artifacts, and electromyographic interference—posing a major challenge in clinical settings. This paper presents a lightweight deep learning-based denoising framework, forming a compact autoencoder architecture. The model was trained under severe noise conditions (−5 dB signal-to-noise ratio (SNR)) using a rigorously partitioned dataset to ensure no data leakage and robust generalization.

Extensive evaluations were conducted across seven noise configurations and three SNR levels (−5 dB, 0 dB, and +5 dB), showing consistent denoising performance with minimal morphological distortion—critical for maintaining diagnostic integrity. In particular, tests on clinically vital rhythms such as ventricular tachycardia (VT) and ventricular fibrillation (VF) confirm that the proposed model effectively suppresses noise without altering arrhythmic features essential for diagnosis. Visual and quantitative assessments, including SNR improvement, RMSE, and correlation metrics, validate the model's efficacy in preserving waveform fidelity.

To demonstrate real-world applicability, the model was deployed on a Raspberry Pi 4 using TensorFlow Lite with float16 precision. Inference latency was measured at just 1.41 seconds per 14-second ECG segment, indicating feasibility for near-real-time use in edge devices. Overall, this study introduces a lightweight, hardware-validated, and morphologically reliable ECG denoising solution suitable for integration into portable or wearable healthcare systems.

# Introduction

ECG plays a fundamental role in cardiovascular diagnostics, providing essential information about heart rhythms, electrical conduction, and conditions such as arrhythmias, heart attacks, and conduction abnormalities[1][2]. However, in real-world applications recorded ECG particularly in ambulatory monitoring and telemedicine are often affected by noise artifacts that can mask important diagnostic features[3]. These include baseline wander (BW) caused by respiration, motion artifacts (MA) resulting from patient movement or loose electrodes, and electromyographic (EMG) noise originating from muscle activity[4]. Effectively removing these noise sources without altering the essential morphology of the ECG signal remains a significant and ongoing challenge[4][5][6].

Traditional ECG denoising techniques—such as wavelet transforms, adaptive filtering, and empirical mode decomposition (EMD)—have been widely used due to their conceptual simplicity and moderate success in separating noise from signal. Wavelet-based methods, for instance, are effective at isolating components in both time and frequency domains and have been employed to filter out baseline drift or high-frequency noise. Adaptive filters, on the other hand, can dynamically adjust to varying noise conditions, while EMD decomposes signals into intrinsic mode functions to isolate noise-dominant components[7][8][9].

Despite their utility, these classical methods face notable limitations when noise shares spectral or morphological characteristics with important ECG features. For example, motion artifacts and muscle noise often overlap with the frequency bands of P-waves, T-waves, and QRS complexes, leading to signal distortion or loss of clinically relevant information[4]. These overlaps make it difficult for handcrafted filters to distinguish between noise and meaningful signal content, especially under low SNR conditions.[10]

To overcome these challenges, there has been a growing shift toward data-driven approaches, particularly deep learning[11][12]. Unlike traditional methods, deep learning models do not rely on predefined rules or assumptions about the signal. Instead, they learn hierarchical feature representations from data itself, enabling more nuanced differentiation between signal and noise. These models have demonstrated improved generalization across different noise types, better preservation of ECG morphology, and enhanced robustness in challenging scenarios such as ambulatory monitoring or cardiopulmonary resuscitation (CPR) conditions[13][14][15][16].

A range of deep learning models including convolutional neural networks, recurrent networks, and adversarial frameworks have been explored for ECG denoising. In [17] the authors proposed a model that combines wavelet transforms with an autoencoder to reduce residual noise left after initial filtering. In [18] authors introduced the Adversarial Denoising Convolutional Neural Network (ADnCNN), which integrates a discriminator to enhance denoising performance. In [19] authors applied Conditional Generative Adversarial Networks (CGANs) to this task, enabling the model to address both single and composite noise scenarios through conditional training.

Despite these advances, many models either fail to explicitly address generalization across noise configurations or lack rigorous data partitioning protocols, often mixing training and testing signals or noise sources. In contrast, this work focuses on improving generalizability by training under severe noise conditions (−5 dB SNR) using a carefully curated dataset, with strict subject-wise and noise-level partitioning to eliminate any risk of data leakage so that independent test data and noise sources are used. Additionally, the model's performance is evaluated not only under normal arrhythmia conditions but also on life-threatening arrhythmic patterns—such as rapid ventricular tachycardia (rapid VT) and coarse ventricular fibrillation (coarse VF)—from the from a real-world AED dataset.

In summary, the contributions of this paper are:

- A light-weighted deep convolutional-recurrent autoencoder with skip connections for robust ECG denoising under -5 dB SNR.
- A rigorously partitioned training and testing pipeline
- Comprehensive evaluation across seven noise configurations and critical arrhythmia types, including generalization to out-of-distribution test data.
- Deployment performance benchmarking on a Raspberry Pi 4 to assess feasibility for edge-device use near-real-time medical settings.

## 3. Method

### 3.1 Overview

We propose a deep-learning-based ECG denoising method using a custom autoencoder model enhanced with bidirectional recurrent layers and skip connections. The model is trained under severe noise conditions of -5 dB SNR using carefully structured training and testing datasets that eliminate any form of data leakage such that data and noise sources independence is established from the training datasets. Unlike prior studies that often lack transparency in how noisy ECGs are generated, we implement a clearly partitioned, reproducible pipeline to ensure methodological integrity and model generalization.

### 3.2 Data Sources and Preprocessing

We used two data sources:

- Clean ECGs: Drawn from the MIT-BIH Arrhythmia Database and a set of manually extracted shockable rhythm segments. All signals were segmented into 14-second windows
- Noise signals: Extracted from the MIT-BIH Noise Stress Test Database, including Electromyographic (EMG) noise, Baseline Wander (BW), and Motion Artifact (MA).

Each noise type consisted of 128 total segments, of which 103 were reserved for training and 25 for testing. Clean ECG data was similarly organized:

- Training set: 4,864 clean segments from MIT-BIH subjects, along with 609 shockable rhythm segments — totaling 5,473 clean ECG segments for training.
- Test set: 1,280 clean ECG segments reserved from unseen MIT-BIH subjects. Additionally, a test set for VF and rapid VT from real-world AED dataset including 41 coarse VF and 31 rapid VT

To synthesize noisy signals, clean segments were combined with scaled versions of EMG, BW, and MA noise. Each noise source contributed equally to the total noise power to meet a -5 dB composite SNR target. A 4th-order Butterworth bandpass filter (0.5–45 Hz) was applied to the noisy signal post-synthesis to remove irrelevant frequency components while preserving diagnostic features.

## 3.3 Strict Dataset Splitting to Maintain Independent Training and Test Datasets

To maintain dataset integrity, we applied a two-level split strategy:

1. Subject-wise data split: Clean ECGs from the MIT-BIH dataset were divided by subject into 80% for training and 20% for testing, ensuring no subject overlap between these sets.
2. Noise-wise split: Each noise type was split into 80% training (103 segments) and 20% testing (25 segments) prior to use.

These splits were executed before any noise synthesis or data augmentation. As such, all training combinations use exclusively training-clean ECGs and training-noise segments, while the test set uses only unseen ECGs and noise segments. This careful design eliminates any form of data leakage, a problem frequently unaddressed or vaguely handled in related works.

## 3.4 Autoencoder Architecture

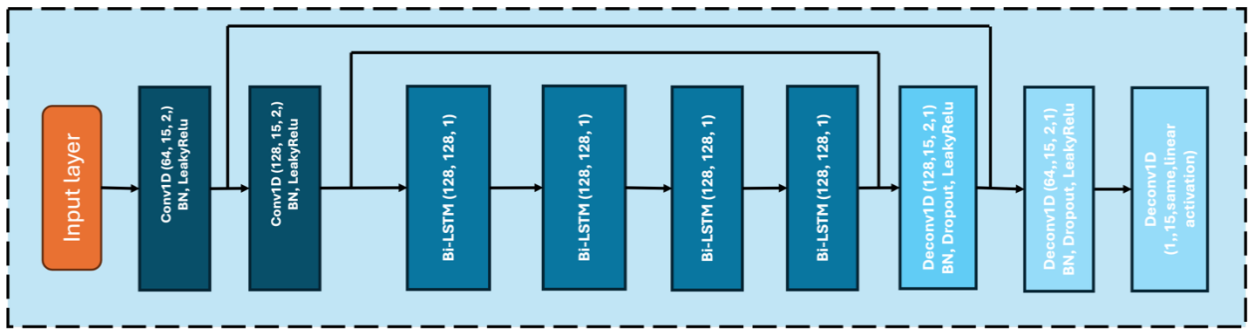

Figure1. Illustration of the model architecture

The proposed model is a deep convolutional autoencoder integrated with Bi-LSTM layers and skip connections. The encoder consists of only two 1D convolutional layers and two Bi-LSTM layers, mirrored symmetrically in the decoder.

*Encoder*

The encoder consists of two 1D convolutional layers (kernel size = 15, stride = 2), each followed by batch normalization and LeakyReLU activation. These are followed by two stacked Bi-LSTM layers capturing forward and backward temporal dynamics of the ECG signal.

*Decoder*

The decoder is a symmetric inverse of the encoder, consisting of two Bi-LSTM layers followed by transposed 1D convolutional layers (Conv1DTranspose). Skip connections are introduced from encoder to decoder to preserve high-resolution features.

## 3.5 Training Setup

The total training procedure involves synthesizing 563,719 noisy ECG segments by combining each of the 5,473 clean training segments with triplets (either separate BW, MA and EMG or the combination of these artifacts ) of 103 training noise segments. This approach results in diverse noisy signals under the same -5 dB SNR constraint, aiding generalization.

The model was trained for 6 epochs using the Adam optimizer (learning rate = 0.001) with mean squared error (MSE) as the loss function. A validation set of 10,000 noisy-clean ECG pairs was prepared using the held-out validation portion of both ECG and noise data. This validation set was used to optimize the network's hyperparameters and monitor training performance. It is important to note that the test set and evaluation protocol differ from those used during validation.

## 3.6 Evaluation Protocol

We evaluated the proposed model's denoising performance using a rigorously structured test framework designed to measure both generalizability and clinical robustness under realistic noise conditions.

*Test Data Construction*

To ensure complete independence from training and validation phases, we synthesized the test dataset using:

- 1,280 clean ECG segments from previously unseen MIT-BIH subjects.
- 25 test segments each of EMG, BW, and MA noises from the held-out portion of the MIT-BIH Noise Stress Test Database.

Each clean segment was combined with scaled contributions from all three noise types, following the same composite noise generation strategy used during training, targeting a $-5$ dB SNR. A 4th-order Butterworth bandpass filter (0.5–45 Hz) was applied before feeding to the network to ensure signal fidelity and diagnostic relevance.

*Evaluation Metrics*

We assessed denoising quality using both quantitative metrics and qualitative visualization:

- Signal-to-Noise Ratio (SNR): Measures the ratio of the power of the clean signal to the power of the residual noise after denoising.

- Mean Squared Error (MSE)**:** Measures average squared deviation from the clean signal.
- Root Mean Squared Error (RMSE)**:** Provides an interpretable error magnitude in the signal's native scale.
- Pearson Correlation Coefficient (r)**:** Quantifies morphological similarity between the clean and denoised waveforms.

*Test Scenarios*

To comprehensively evaluate model robustness, we analyzed performance under:

- Seven noise conditions: Each individual noise type and all pairwise and combined scenarios.
- Three SNR levels: −5 dB, 0 dB, and +5 dB to test noise-level generalization.
- Subject-wise segmentation: To assess inter-subject generalization.
- Shockable rhythms (Rapid VT, Coarse VF): Using manually extracted from an AED.

*Visual Inspection*

We conducted waveform-level analysis by plotting clean, noisy, and denoised signals for selected arrhythmias (e.g., AF, PVC, PAC). This visual assessment ensures the model preserves diagnostic features post-denoising.

## 3.7 Hardware Evaluation

To evaluate the real-time feasibility of the proposed ECG denoising framework, we deployed the trained model on a Raspberry Pi 4 Model B—an affordable and widely accessible single-board computer. This deployment is particularly relevant for use cases involving wearable medical devices or portable health-monitoring systems, where resource constraints are a key concern.

The Raspberry Pi 4 used for deployment has the following specifications:

- CPU: ARM Cortex-A72 (quad-core, 64-bit, 1.5 GHz)
- RAM: 8 GB LPDDR4
- Architecture: aarch64 (64-bit)

The device was evaluated in a headless configuration using Ubuntu 64-bit and standard Python libraries. Model inference was executed using TensorFlow Lite with optimizations for reduced precision (float16), improving compatibility with edge devices.

## 4.Results

We evaluated the proposed ECG denoising method using a test set composed entirely of ECG segments from independent subjects, distinct from those used during training. Clean ECG signals were drawn from the test partition of the MIT-BIH dataset and combined with noise samples from a separately held-out noise test set. This strict separation of subjects and noise sources ensures that the evaluation is completely free of data leakage.

The first set of tables summarizes the model's denoising performance across seven noise configurations (EMG, baseline wander, motion artifact, and their combinations) under three SNR levels: −5 dB, 0 dB, and +5 dB. In addition, we provide subject-wise results to evaluate performance consistency across individuals. The evaluation also includes results on coarse VF and rapid VT segments, which were sourced from a real-world AED dataset , to assess robustness in life-threatening arrhythmic conditions. Furthermore, the figures illustrate the model's performance on segments containing specific arrhythmias, including atrial fibrillation (AF), premature atrial contractions (PAC), and premature ventricular contractions (PVC). This analysis is particularly important from a clinical perspective, as it demonstrates that the model not only removes noise effectively but also preserves the morphological features critical for arrhythmia interpretation—ensuring that cardiologists can still accurately assess the underlying cardiac events in the denoised signals.

It is important to highlight that the model was trained solely on −5 dB noisy data but evaluated across a range of SNR conditions. The results show that while SNR increases from −5 dB to +5 dB, model performance does not significantly improve, indicating that the model is already highly effective under severe noise. As illustrated in the figure 2 to 11, the denoising capability at −5 dB is strong enough to reach a performance plateau, suggesting saturation beyond this level. Note that in these tables high correlation values were obtained between the clean and denoised signals for all SNR conditions. Even in the cases involving VF and VT, relatively high correlation values are obtained despite various scenarios of artifact contamination with SNR at -5 dB, as shown in Table 5. Figures 2-11 show representative plots of denoising capability of the proposed deep learning architecture on various types of arrhythmias including: normal sinus rhythm, AF, PAC, PVC, VT and VF. Note the similarity in waveform morphologies between the clean and denoised signals in these figures. Quantitative assessment of waveform morphologies can be inferred from correlation values provided in Tables 1-5.

For the hardware implementation of the algorithm on Raspberry Pi 4 Model B, the average latency per inference on 14-second ECG segments (5,040 samples) was measured at 1.41 seconds, including preprocessing and postprocessing steps. This latency demonstrates that the model is suitable for near real-time applications in mobile or ambulatory environments, such as CPR monitoring or field diagnostics. While not meeting hard real-time constraints, the performance is sufficient for asynchronous analysis or buffered streaming applications in low-resource settings.

| Noise Combination | SNR (dB) | MSE | RMSE | Correlation |
|---|---|---|---|---|
| **EMG** | 6.88 ± 2.71 | 0.0421 ± 0.0482 | 0.1806 ± 0.0975 | 0.8733 ± 0.0836 |
| **BW** | 10.29 ± 2.82 | 0.0228 ± 0.0374 | 0.1255 ± 0.0842 | 0.9491 ± 0.0436 |
| **MA** | 7.99 ± 2.68 | 0.0340 ± 0.0447 | 0.1594 ± 0.0927 | 0.9060 ± 0.0689 |
| **EMG + BW** | 8.24 ± 2.59 | 0.0317 ± 0.0394 | 0.1549 ± 0.0876 | 0.9120 ± 0.0596 |
| **EMG + MA** | 7.07 ± 2.70 | 0.0408 ± 0.0495 | 0.1767 ± 0.0978 | 0.8786 ± 0.0849 |
| **BW + MA** | 8.94 ± 2.58 | 0.0283 ± 0.0400 | 0.1437 ± 0.0875 | 0.9282 ± 0.0522 |
| **EMG + BW + MA** | 7.93 ± 2.57 | 0.0338 ± 0.0432 | 0.1599 ± 0.0906 | 0.9050 ± 0.0645 |
| **Overall** | 8.19 ± 1.07 | 0.0334 ± 0.0062 | 0.1572 ± 0.0175 | 0.9075 ± 0.0245 |

Table 1) Overall Denoising Performance by Noise Type at -5 dB SNR

| Subject | SNR (dB) ± SD | MSE ± SD | RMSE ± SD | Correlation ± SD |
|---|---|---|---|---|
| **103** | 12.72 ± 0.83 | 0.0055 ± 0.0010 | 0.0717 ± 0.0066 | 0.9703 ± 0.0056 |
| **105** | 8.98 ± 0.62 | 0.0390 ± 0.0030 | 0.1473 ± 0.0078 | 0.9074 ± 0.0109 |
| **111** | 7.20 ± 1.05 | 0.0081 ± 0.0018 | 0.0863 ± 0.0100 | 0.8933 ± 0.0298 |
| **116** | 6.12 ± 1.11 | 0.0788 ± 0.0160 | 0.2688 ± 0.0298 | 0.8757 ± 0.0440 |
| **122** | 7.10 ± 2.17 | 0.0309 ± 0.0138 | 0.1672 ± 0.0400 | 0.8755 ± 0.0569 |
| **205** | 8.85 ± 0.89 | 0.0049 ± 0.0011 | 0.0673 ± 0.0071 | 0.9276 ± 0.0156 |
| **213** | 6.79 ± 0.31 | 0.0854 ± 0.0062 | 0.2910 ± 0.0104 | 0.8985 ± 0.0095 |
| **219** | 8.42 ± 2.20 | 0.0327 ± 0.0130 | 0.1736 ± 0.0388 | 0.9099 ± 0.0369 |
| **223** | 8.85 ± 1.72 | 0.0238 ± 0.0090 | 0.1460 ± 0.0285 | 0.9170 ± 0.0365 |
| **230** | 6.90 ± 0.16 | 0.0246 ± 0.0009 | 0.1533 ± 0.0027 | 0.8993 ± 0.0060 |

Table 2) Subject-Wise Average Denoising Performance Across All Noise Types at -5 dB SNR

| Noise Combination | SNR (dB) ± SD | MSE ± SD | RMSE ± SD | Correlation ± SD |
|---|---|---|---|---|
| **EMG** | 9.32 ± 2.41 | 0.0241 ± 0.0319 | 0.1350 ± 0.0766 | 0.9368 ± 0.0402 |
| **BW** | 10.91 ± 2.71 | 0.0188 ± 0.0291 | 0.1151 ± 0.0742 | 0.9581 ± 0.0330 |
| **MA** | 9.72 ± 2.49 | 0.0230 ± 0.0315 | 0.1302 ± 0.0778 | 0.9432 ± 0.0392 |
| **EMG + BW** | 9.89 ± 2.43 | 0.0215 ± 0.0301 | 0.1268 ± 0.0738 | 0.9460 ± 0.0350 |
| **EMG + MA** | 9.33 ± 2.38 | 0.0242 ± 0.0331 | 0.1349 ± 0.0774 | 0.9374 ± 0.0409 |
| **BW + MA** | 10.19 ± 2.53 | 0.0211 ± 0.0299 | 0.1239 ± 0.0758 | 0.9500 ± 0.0355 |
| **EMG + BW + MA** | 9.70 ± 2.41 | 0.0226 ± 0.0316 | 0.1297 ± 0.0757 | 0.9432 ± 0.0371 |
| **Overall** | 9.87 ± 0.51 | 0.0222 ± 0.0018 | 0.1279 ± 0.0064 | 0.9449 ± 0.0068 |

Table 3) Overall Denoising Performance by Noise Type at 0 dB SNR

| Subject | SNR (dB) ± SD | MSE ± SD | RMSE ± SD | Correlation ± SD |
|---|---|---|---|---|
| **103** | 13.80 ± 0.42 | 0.0042 ± 0.0004 | 0.0628 ± 0.0029 | 0.9799 ± 0.0020 |
| 105 | 10.02 ± 0.30 | 0.0294 ± 0.0011 | 0.1297 ± 0.0033 | 0.9305 ± 0.0038 |
| 111 | 8.14 ± 0.41 | 0.0061 ± 0.0006 | 0.0763 ± 0.0036 | 0.9270 ± 0.0108 |
| 116 | 8.38 ± 0.57 | 0.0495 ± 0.0053 | 0.2108 ± 0.0120 | 0.9424 ± 0.0148 |
| 122 | 9.75 ± 1.06 | 0.0146 ± 0.0033 | 0.1188 ± 0.0140 | 0.9497 ± 0.0092 |
| 205 | 10.12 ± 0.52 | 0.0034 ± 0.0004 | 0.0572 ± 0.0034 | 0.9541 ± 0.0063 |
| 213 | 7.90 ± 0.15 | 0.0661 ± 0.0023 | 0.2561 ± 0.0045 | 0.9276 ± 0.0035 |
| 219 | 10.79 ± 1.14 | 0.0175 ± 0.0044 | 0.1291 ± 0.0167 | 0.9533 ± 0.0123 |
| 223 | 11.81 ± 0.68 | 0.0116 ± 0.0016 | 0.1028 ± 0.0077 | 0.9646 ± 0.0071 |
| 230 | 7.94 ± 0.11 | 0.0192 ± 0.0005 | 0.1358 ± 0.0017 | 0.9204 ± 0.0022 |

Table 4) Subject-Wise Average Denoising Performance Across All Noise Types at 0 dB SNR

| Noise Combination | SNR (dB) ± SD | MSE ± SD | RMSE ± SD | Correlation ± SD |
|---|---|---|---|---|
| EMG | 10.22 ± 2.49 | 0.0203 ± 0.0291 | 0.1225 ± 0.0729 | 0.9503 ± 0.0326 |
| BW | 10.97 ± 2.73 | 0.0186 ± 0.0290 | 0.1145 ± 0.0743 | 0.9588 ± 0.0330 |
| MA | 10.44 ± 2.58 | 0.0202 ± 0.0290 | 0.1207 ± 0.0751 | 0.9531 ± 0.0344 |
| EMG + BW | 10.49 ± 2.57 | 0.0196 ± 0.0292 | 0.1195 ± 0.0732 | 0.9531 ± 0.0327 |
| EMG + MA | 10.21 ± 2.50 | 0.0206 ± 0.0297 | 0.1230 ± 0.0741 | 0.9500 ± 0.0340 |
| BW + MA | 10.65 ± 2.63 | 0.0195 ± 0.0284 | 0.1180 ± 0.0743 | 0.9556 ± 0.0336 |
| EMG + BW + MA | 10.38 ± 2.54 | 0.0201 ± 0.0295 | 0.1209 ± 0.0740 | 0.9521 ± 0.0336 |
| Overall | 10.48 ± 0.25 | 0.0198 ± 0.0006 | 0.1199 ± 0.0027 | 0.9533 ± 0.0028 |

Table 5) Overall Denoising Performance by Noise Type at +5 dB SNR

| Subject | SNR (dB) ± SD | MSE ± SD | RMSE ± SD | Correlation ± SD |
|---|---|---|---|---|
| 103 | 14.34 ± 0.19 | 0.0037 ± 0.0001 | 0.0590 ± 0.0012 | 0.9825 ± 0.0007 |
| 105 | 10.38 ± 0.14 | 0.0280 ± 0.0005 | 0.1254 ± 0.0013 | 0.9355 ± 0.0014 |
| 111 | 8.57 ± 0.22 | 0.0054 ± 0.0003 | 0.0725 ± 0.0018 | 0.9385 ± 0.0054 |
| 116 | 9.18 ± 0.21 | 0.0420 ± 0.0016 | 0.1938 ± 0.0039 | 0.9610 ± 0.0051 |
| 122 | 10.84 ± 0.71 | 0.0110 ± 0.0018 | 0.1039 ± 0.0085 | 0.9606 ± 0.0050 |
| 205 | 10.71 ± 0.24 | 0.0029 ± 0.0002 | 0.0532 ± 0.0015 | 0.9616 ± 0.0022 |
| 213 | 8.00 ± 0.07 | 0.0644 ± 0.0010 | 0.2528 ± 0.0021 | 0.9310 ± 0.0017 |
| 219 | 12.07 ± 0.55 | 0.0125 ± 0.0016 | 0.1102 ± 0.0071 | 0.9670 ± 0.0048 |
| 223 | 12.70 ± 0.28 | 0.0095 ± 0.0005 | 0.0927 ± 0.0028 | 0.9729 ± 0.0031 |
| 230 | 8.01 ± 0.06 | 0.0191 ± 0.0002 | 0.1350 ± 0.0009 | 0.9221 ± 0.0012 |

Table 6) Subject-Wise Average Denoising Performance Across All Noise Types at 5 dB SNR

| ECG Type | Noise Type | SNR (dB) ± SD | MSE ± SD | RMSE ± SD | Correlation ± SD |
|---|---|---|---|---|---|
| **Coarse VF** | EMG | 4.64 ± 4.08 | 0.1422 ± 0.3160 | 0.2835 ± 0.2487 | 0.7317 ± 0.2731 |
| **Coarse VF** | BW | 9.56 ± 3.65 | 0.0509 ± 0.1096 | 0.1660 ± 0.1529 | 0.9212 ± 0.0896 |
| **Coarse VF** | MA | 6.57 ± 4.63 | 0.0926 ± 0.1902 | 0.2294 ± 0.1999 | 0.8064 ± 0.2531 |
| **Coarse VF** | EMG+BW | 5.94 ± 4.09 | 0.1052 ± 0.2273 | 0.2439 ± 0.2139 | 0.7740 ± 0.2630 |
| **Coarse VF** | EMG+MA | 5.04 ± 3.94 | 0.1297 ± 0.2961 | 0.2693 ± 0.2392 | 0.7421 ± 0.2778 |
| **Coarse VF** | BW+MA | 7.42 ± 4.51 | 0.0819 ± 0.1836 | 0.2104 ± 0.1940 | 0.8423 ± 0.2161 |
| **Coarse VF** | EMG+BW+MA | 5.77 ± 3.90 | 0.1127 ± 0.2581 | 0.2490 ± 0.2251 | 0.7730 ± 0.2589 |
| **Rapid VT** | EMG | 6.14 ± 4.02 | 0.2277 ± 0.4024 | 0.3815 ± 0.2867 | 0.7985 ± 0.2358 |
| **Rapid VT** | BW | 11.21 ± 4.20 | 0.0882 ± 0.1764 | 0.2276 ± 0.1908 | 0.9402 ± 0.0725 |
| **Rapid VT** | MA | 8.31 ± 3.68 | 0.1284 ± 0.2025 | 0.2954 ± 0.2027 | 0.8770 ± 0.1536 |
| **Rapid VT** | EMG+BW | 7.87 ± 3.88 | 0.1590 ± 0.2784 | 0.3165 ± 0.2426 | 0.8631 ± 0.1797 |
| **Rapid VT** | EMG+MA | 6.76 ± 3.67 | 0.1873 ± 0.3075 | 0.3515 ± 0.2526 | 0.8361 ± 0.1960 |
| **Rapid VT** | BW+MA | 9.17 ± 3.76 | 0.1130 ± 0.1962 | 0.2716 ± 0.1980 | 0.9036 ± 0.1103 |
| **Rapid VT** | EMG+BW+MA | 7.58 ± 3.60 | 0.1527 ± 0.2412 | 0.3199 ± 0.2243 | 0.8612 ± 0.1770 |

Table 5) Denoising Performance on VF and VT at −5 dB SNR Across Different Noise Combination

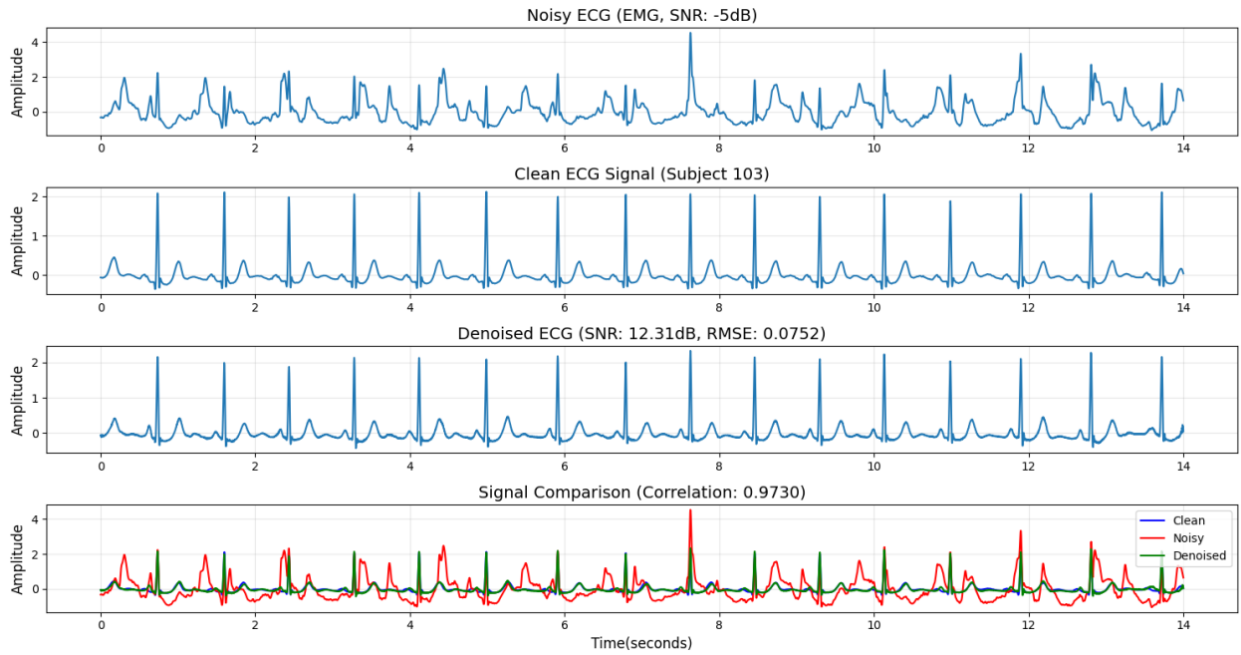

Figure 2) Illustration of denoising performance on an ECG signal corrupted by EMG noise

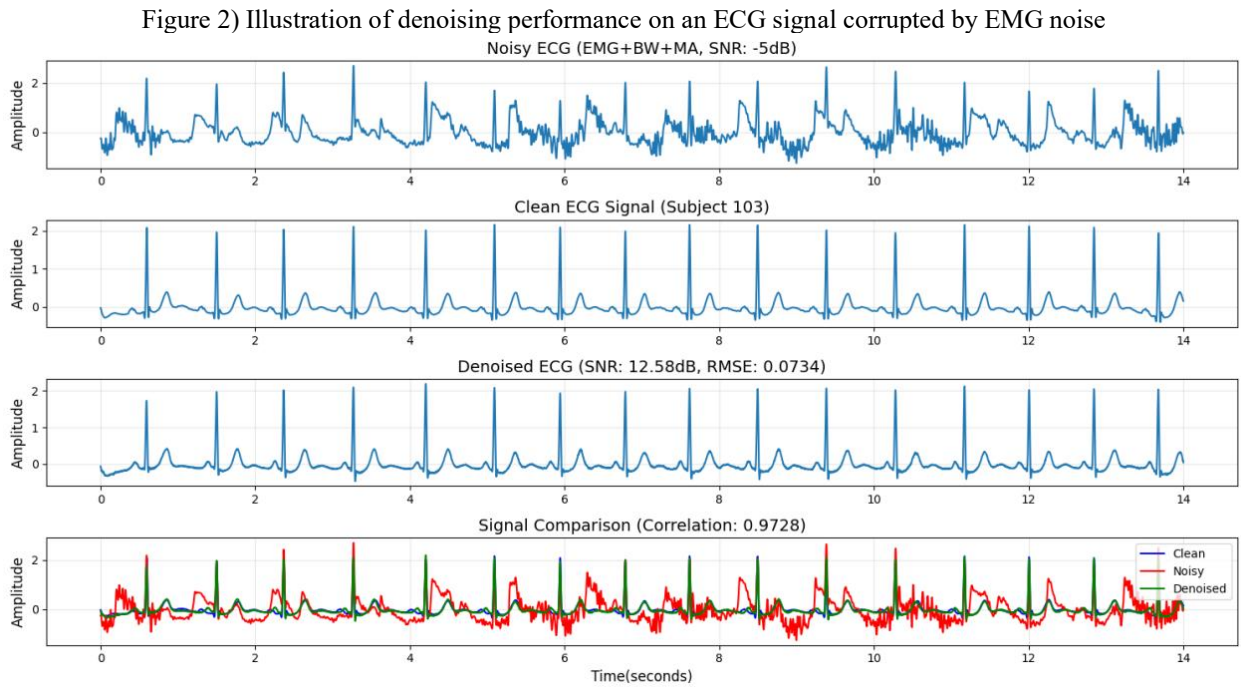

Figure 3) Illustration of denoising performance on an ECG signal corrupted by combination of EMG+BW+MA noise

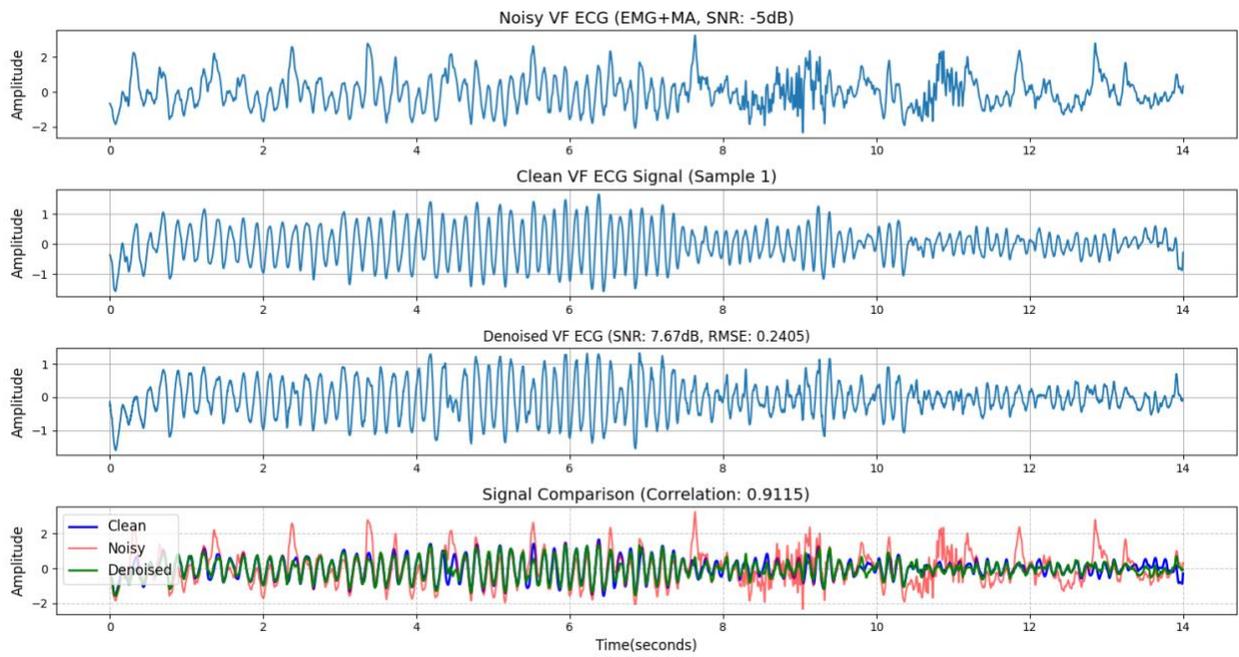

Figure 4) Illustration of denoising performance on a coarse VF ECG corrupted by combination of EMG+MA noise

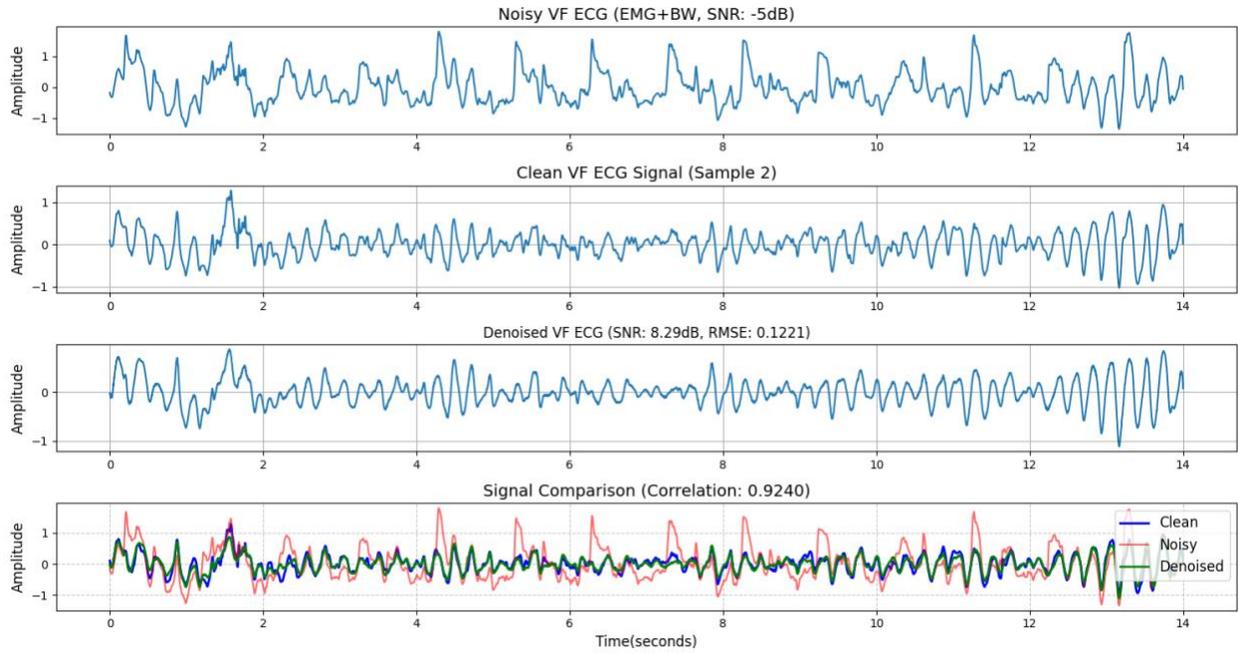

figure 5 ) Illustration of denoising performance on coarse VF ECG corrupted by combination of EMG+MA noise

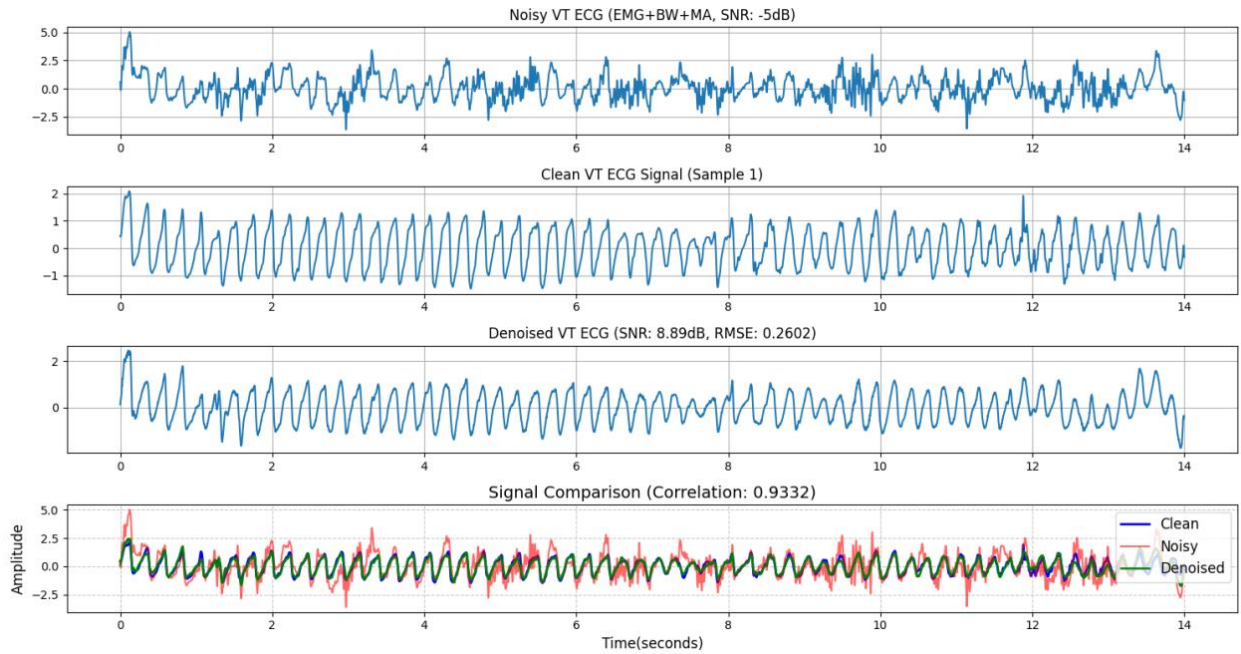

Figure 6) Illustration of denoising performance on a rapid VT ECG corrupted by combination of EMG+BW+MA noise

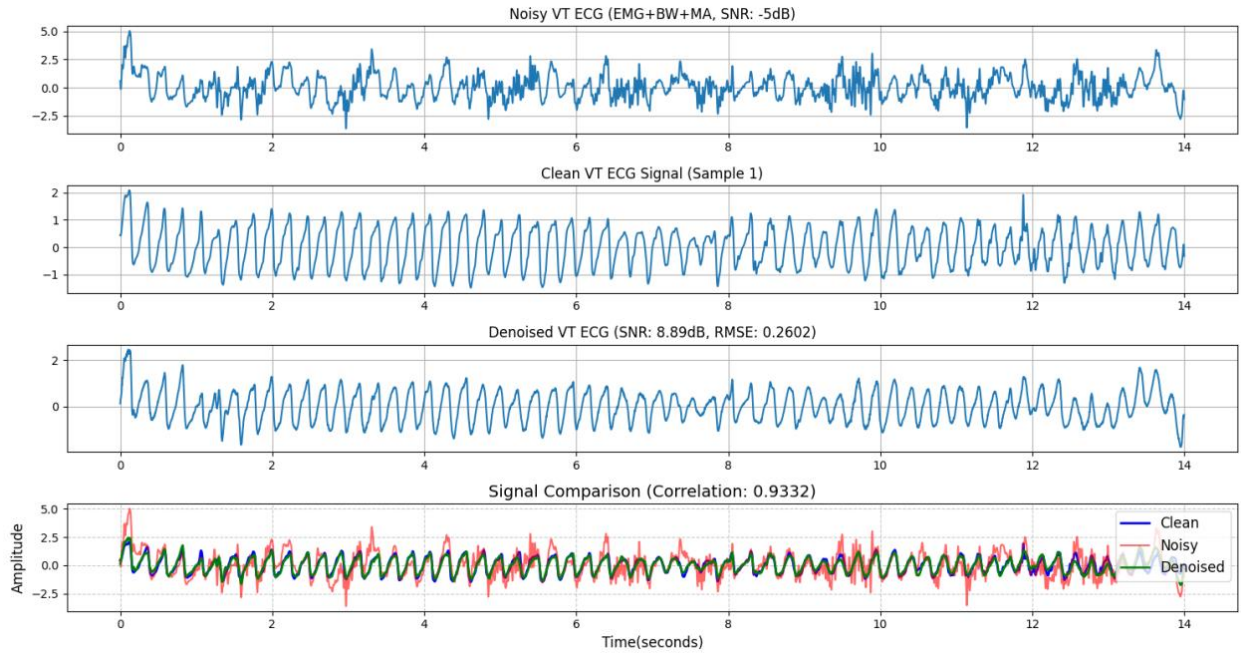

Figure 7) Illustration of denoising performance on a rapid VT ECG corrupted by combination of EMG+BW+MA noise

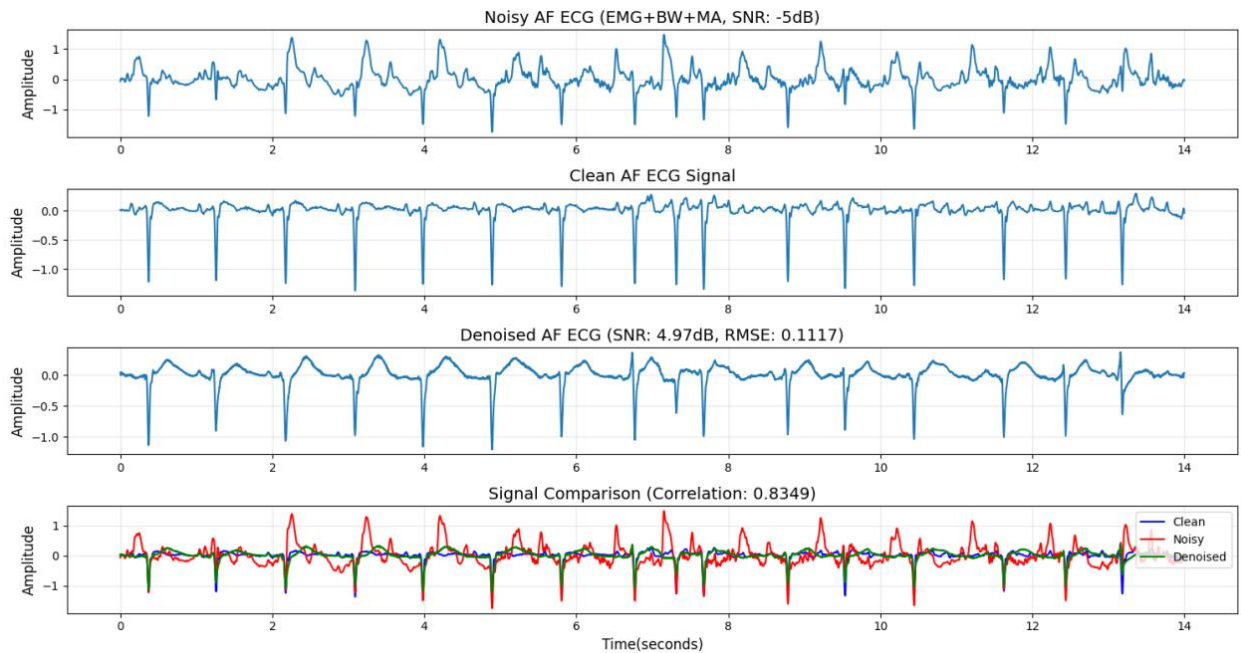

Figure 8) Illustration of denoising performance on an AF segment corrupted by combination of EMG+BW+MA noise

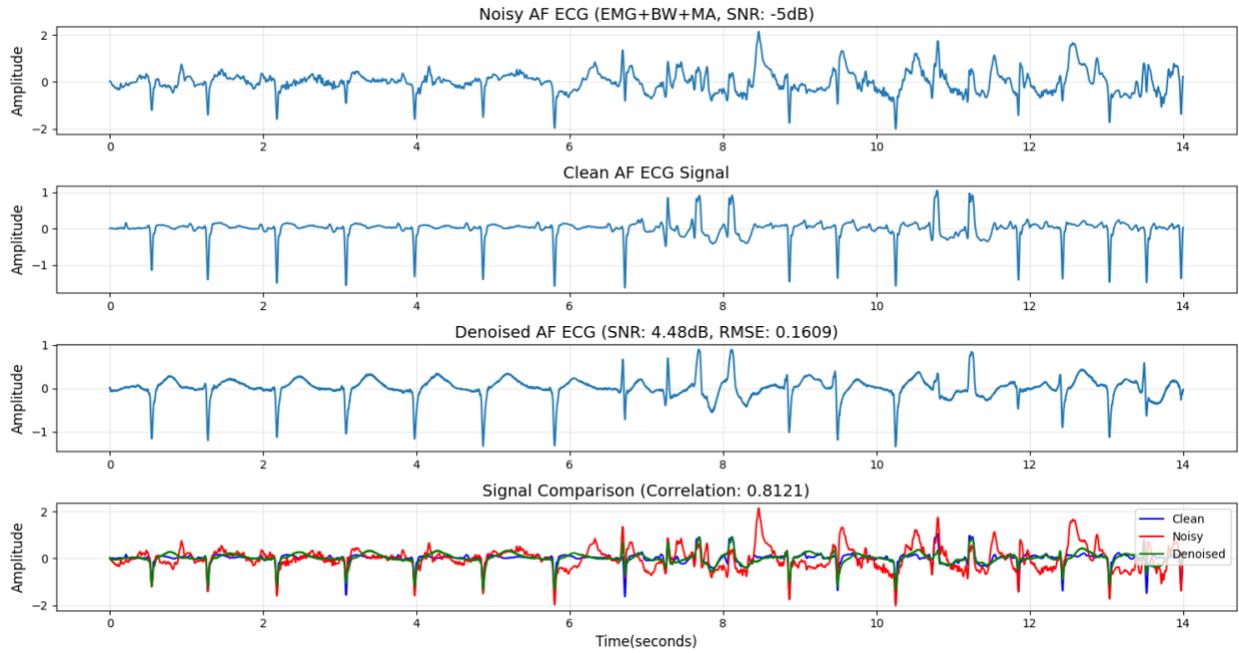

Figure 9) Illustration of denoising performance on an AF and PVC segment corrupted by combination of EMG+BW+MA noise

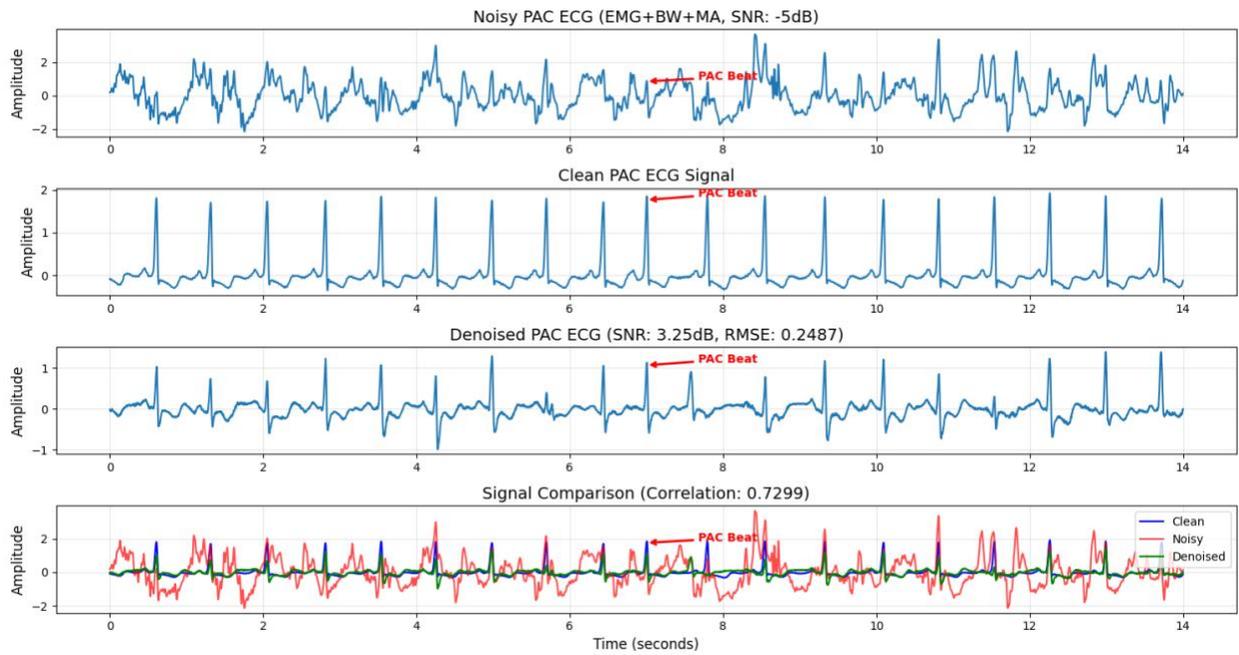

Figure 10) Illustration of denoising performance on ECG segment containing PAC beat corrupted by combination of EMG+BW+MA noise

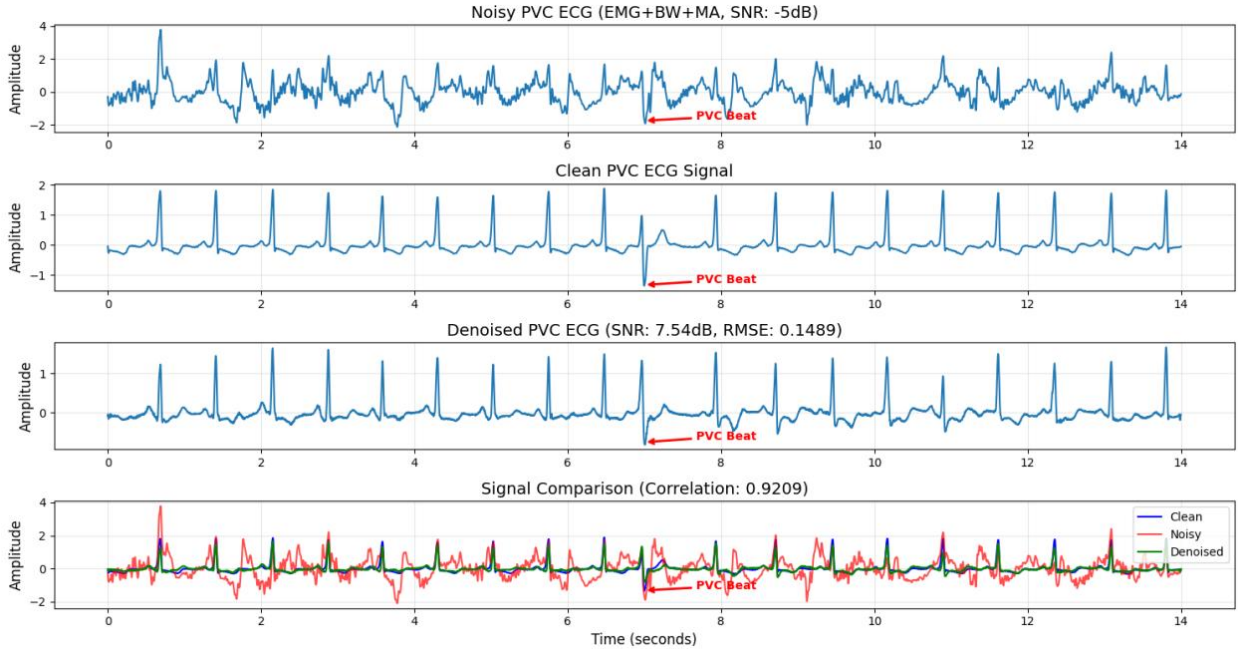

Figure 11) Illustration of denoising performance on ECG segment containing PVC beat corrupted by combination of EMG+BW+MA noise

## 5.Discussion

While numerous recent studies have explored deep learning methods for ECG denoising, many fall short in ensuring experimental rigor especially involving subject and noise source independence amongst training and test datasets. In our work, we address these critical gaps by designing a lightweight neural network tailored for arrhythmia-specific denoising, leveraging a rich and diverse dataset, and emphasizing strict data integrity throughout the pipeline.

One of the key strengths of our approach lies in the explicit separation of clean ECG and noise signals prior to their combination for training and testing. Unlike several previous studies where the same or similar noise segments could inadvertently appear in both training and testing sets—leading to potential data leakage and inflated performance—our methodology ensures that noise and ECG sources are mutually exclusive across the data splits. This careful design significantly improves the generalizability and trustworthiness of the reported performance, especially under real-world noisy conditions.

Moreover, we evaluate the model on arrhythmia-specific cases, such as rapid VT and coarse VF, which are both clinically critical and morphologically distinct. We also provide visual comparisons of model performance across different arrhythmias, highlighting how clean ECG morphology affects denoising outcomes—something rarely addressed in prior works. This adds a valuable layer of interpretability and clinical relevance to our findings.

In terms of data diversity, our work utilizes a comprehensive combination of clean ECG signals from the MIT-BIH Arrhythmia Database and noise types from the NSTDB, including muscle noise, baseline wander, and electrode motion artifact. By generating all possible noise combinations, we simulate a wide range of realistic and challenging scenarios, providing a robust benchmark for ECG denoising. This stands in contrast to many existing studies that either use limited or poorly documented noise configurations.

Finally, to assess the practical applicability of our model, we implement it on a Raspberry Pi 4, a widely accessible edge-computing device. It achieves a reasonable latency of 1.41 seconds, demonstrating the potential for real-time deployment in low-resource or wearable settings.

In contrast to our approach, it is important to critically evaluate how other recent works have addressed similar challenges in ECG denoising. While some studies claim impressive performance, a closer inspection reveals limitations in terms of evaluation rigor, reproducibility, and real-world applicability.

In [19] while the proposed ECG denoising method using CAE-CGAN reports promising results, it also presents several notable limitations. Most strikingly, the reported signal-to-noise ratio (SNR) improvements—approaching 40 dB across various noise conditions—are exceptionally high, particularly under severe scenarios such as 0 dB input SNR. These results raise concerns regarding potential overfitting or an evaluation protocol that may not accurately capture the variability and complexity of real-world ECG signals. Furthermore, the paper provides insufficient detail about its noise contamination procedure. Without a transparent and rigorously defined noise contamination process, the reproducibility and generalizability of the results remain uncertain.

In [17] authors presented a denoising method based on a combination of wavelet transform and a denoising autoencoder, achieving strong performance. However, the study consistently trains on a fixed set of 30,000 samples, with only 2,000 reserved for testing. This limited partitioning restricts data diversity and raises concerns about overfitting and generalization. The data augmentation strategy is poorly described and appears inadequate. Because similar noise types may appear in both training and testing sets because of the unclear way of augmentation and splitting , the model may end up memorizing specific noise patterns instead of learning to generalize to new, unseen conditions. The paper fails to clearly state whether noise was added before or after sample selection, introducing ambiguity into the experimental design. In contrast, our method uses a significantly larger and more diverse dataset that spans all possible noise combinations and strictly separates noise and ECG sources between training and testing sets— offering a more rigorous and realistic evaluation framework.

In [18] authors also emphasize separation of ECG recordings between training and testing, which is an important step toward mitigating data leakage. However, the methodology for noise incorporation remains vague. It is not clear whether noise was added before or after data splitting, or whether the same types or segments of noise could have been used across both datasets. This lack of clarity is critical, particularly for denoising tasks, where the presence of similar noise patterns in both sets can lead to overfitting and exaggerated performance. Robust denoising models must ensure not only that clean ECG signals are kept distinct but also that noise segments are independently distributed across the splits. Without this, the validity of the evaluation becomes questionable.

In [20] authors demonstrate strong performance in only removing baseline wander. Their handling of noise remains ambiguous. While the authors ensure that ECG signals come from different subjects for training and testing, they provide insufficient detail on how noise from the MIT-BIH Noise Stress Test Database was introduced. It remains unclear whether noise segments were reused across datasets or randomly distributed. This omission is important, as repeated exposure to the same noise in both training and testing can lead to overfitting and inflated results. A thorough evaluation must account for both signal and noise independence to reliably assess model generalization.

In [19] authors proposed a deep learning-based ECG denoising model that performs well under the tested conditions. However, their evaluation is limited primarily to ECG signals with relatively high SNRs (above +5 dB), restricting insight into model performance under lower, more challenging noise levels common in real-world settings. Additionally, while the authors provide a clear description of their training and validation setup, they do not specify how noise was added or whether the same segments were reused. Without a carefully controlled and documented separation of noise and ECG across datasets, there is a risk of data leakage, which can distort the model's perceived robustness.

These shortcomings highlight the need for more rigorous testing in low-SNR environments and clearer noise handling protocols to assess real-world applicability.

## Conclusion

In this study, we introduced a robust and rigorously evaluated deep-learning-based ECG denoising framework suitable for real-world applications from low to high-SNR environments. The proposed model demonstrates strong denoising capabilities while preserving clinically relevant morphological features. Our experimental setup emphasizes reproducibility and integrity, employing strict subject-wise and noise dataset separation to eliminate data leakage which is an often-overlooked issue in previous works

Comprehensive evaluations across multiple noise types, composite noise scenarios, and SNR levels from −5 dB to +5 dB confirm that the model generalizes well under severely to moderately degraded signal conditions. Notably, even when trained solely on −5 dB noisy data, the model exhibits consistent performance across higher SNRs, underscoring its robustness. Subject-wise and arrhythmia-specific analyses further highlight the model's capacity to preserve diagnostic fidelity in life-threatening conditions such as coarse VF and rapid VT.

Another contribution of this work lies in its practical deployment on a resource-constrained device, demonstrating the feasibility of near-real-time inference on edge hardware like the Raspberry Pi 4.

Altogether, this study not only advances the state of ECG denoising with a methodologically sound and clinically relevant approach but also sets a high standard for dataset curation, evaluation transparency, and real-world deployment readiness.

## Blogography

[1] E. Silvetti *et al.*, "The pivotal role of ECG in cardiomyopathies," *Front. Cardiovasc. Med.*, vol. 10, Jun. 2023, doi: 10.3389/fcvm.2023.1178163.
[2] T. Anbalagan, M. K. Nath, D. Vijayalakshmi, and A. Anbalagan, "Analysis of various techniques for ECG signal in healthcare, past, present, and future," *Biomed. Eng. Adv.*, vol. 6, p. 100089, Nov. 2023, doi: 10.1016/j.bea.2023.100089.
[3] A. Ghandehari, J. A. Tavares-Negrete, X. Pei, J. Rajendran, S. Chakoma, and R. Esfandyarpour, "Optimizing NFC-Based Wearable Sensors for Arterial Pulse Monitoring: A Comparative Study of Sampling Rates and Machine Learning Models," in *2024 IEEE 20th International Conference on Body Sensor Networks (BSN)*, Oct. 2024, pp. 1–4. doi: 10.1109/BSN63547.2024.10780749.
[4] M. T. Almalchy, V. Ciobanu, and N. Popescu, "Noise Removal from ECG Signal Based on Filtering Techniques," in *2019 22nd International Conference on Control Systems and Computer Science (CSCS)*, May 2019, pp. 176–181. doi: 10.1109/CSCS.2019.00037.
[5] M. Khan, F. Aslam, T. Zaidi, and S. A. Khan, "Wavelet Based ECG Denoising Using Signal-Noise Residue Method," in *2011 5th International Conference on Bioinformatics and Biomedical Engineering*, May 2011, pp. 1–4. doi: 10.1109/icbbe.2011.5780263.
[6] C. Stolojescu, "Estimation of noise in ECG signals using wavelets," in *2008 11th International Conference on Optimization of Electrical and Electronic Equipment*, May 2008, pp. 113–118. doi: 10.1109/OPTIM.2008.4602509.
[7] Md. A. Kabir and C. Shahnaz, "Denoising of ECG signals based on noise reduction algorithms in EMD and wavelet domains," *Biomed. Signal Process. Control*, vol. 7, no. 5, pp. 481–489, Sep. 2012, doi: 10.1016/j.bspc.2011.11.003.
[8] C. Li, Y. Wu, H. Lin, J. Li, F. Zhang, and Y. Yang, "ECG Denoising Method Based on an Improved VMD Algorithm," *IEEE Sens. J.*, vol. 22, no. 23, pp. 22725–22733, Dec. 2022, doi: 10.1109/JSEN.2022.3214239.
[9] M.-B. Hossain, S. K. Bashar, J. Lazaro, N. Reljin, Y. Noh, and K. H. Chon, "A robust ECG denoising technique using variable frequency complex demodulation," *Comput. Methods Programs Biomed.*, vol. 200, p. 105856, Mar. 2021, doi: 10.1016/j.cmpb.2020.105856.
[10] G. D. Clifford, "ECG Statistics, Noise, Artifacts, and Missing Data".
[11] J. Lim, D. Han, M. Pirayesh Shirazi Nejad, and K. H. Chon, "ECG classification via integration of adaptive beat segmentation and relative heart rate with deep learning networks," *Comput. Biol. Med.*, vol. 181, p. 109062, Oct. 2024, doi: 10.1016/j.compbiomed.2024.109062.
[12] X. Liu, H. Wang, Z. Li, and L. Qin, "Deep learning in ECG diagnosis: A review," *Knowl.-Based Syst.*, vol. 227, p. 107187, Sep. 2021, doi: 10.1016/j.knosys.2021.107187.

[13] M. P. S. Nejad, V. Kargin, S. Hajeb-M, D. Hicks, M. Valentine, and K. H. Chon, "Enhancing the accuracy of shock advisory algorithms in automated external defibrillators during ongoing cardiopulmonary resuscitation using a cascade of CNNEDs," *Comput. Biol. Med.*, vol. 172, p. 108180, Apr. 2024, doi: 10.1016/j.compbiomed.2024.108180.

[14] "Enhancing the accuracy of shock advisory algorithms in automated external defibrillators during ongoing cardiopulmonary resuscitation using a deep convolutional Encoder-Decoder filtering model - ScienceDirect." Accessed: Jun. 26, 2025. [Online]. Available: https://www.sciencedirect.com/science/article/abs/pii/S0957417422008284

[15] "Deep Neural Network Approach for Continuous ECG-Based Automated External Defibrillator Shock Advisory System During Cardiopulmonary Resuscitation | Journal of the American Heart Association." Accessed: Jun. 26, 2025. [Online]. Available: https://www.ahajournals.org/doi/full/10.1161/JAHA.120.019065

[16] J.-P. Didon *et al.*, "Clinical performance of AED shock advisory system with integrated Analyze Whilst Compressing algorithm for analysis of the ECG rhythm during out-of-hospital cardiopulmonary resuscitation: A secondary analysis of the DEFI 2022 study," *Resusc. Plus*, vol. 19, p. 100740, Sep. 2024, doi: 10.1016/j.resplu.2024.100740.

[17] P. Xiong, H. Wang, M. Liu, S. Zhou, Z. Hou, and X. Liu, "ECG signal enhancement based on improved denoising auto-encoder," *Eng. Appl. Artif. Intell.*, vol. 52, pp. 194–202, Jun. 2016, doi: 10.1016/j.engappai.2016.02.015.

[18] Y. Hou, R. Liu, M. Shu, and C. Chen, "An ECG denoising method based on adversarial denoising convolutional neural network," *Biomed. Signal Process. Control*, vol. 84, p. 104964, Jul. 2023, doi: 10.1016/j.bspc.2023.104964.

[19] X. Wang *et al.*, "An ECG Signal Denoising Method Using Conditional Generative Adversarial Net," *IEEE J. Biomed. Health Inform.*, vol. 26, no. 7, pp. 2929–2940, Jul. 2022, doi: 10.1109/JBHI.2022.3169325.

[20] F. P. Romero, D. C. Piñol, and C. R. Vázquez-Seisdedos, "DeepFilter: An ECG baseline wander removal filter using deep learning techniques," *Biomed. Signal Process. Control*, vol. 70, p. 102992, Sep. 2021, doi: 10.1016/j.bspc.2021.102992.